\documentclass[%
 reprint,
 amsmath,amssymb,
 aps,
]{revtex4-2}
\usepackage{graphicx}
\usepackage{dcolumn}
\usepackage{bm}
\usepackage{siunitx,svg, float, lipsum, subfig}
\usepackage{hyperref}

\begin{document}
\preprint{APS/123-QED}

\title{Magneto-optical trap performance for high-bandwidth applications}

\author{B. Adams\textsuperscript{1}, S. Kinge\textsuperscript{2}, K. Bongs\textsuperscript{1,3,4}, Y. H. Lien\textsuperscript{1}}

\affiliation{\textsuperscript{1}Midlands Ultracold Atom Research Centre, School of Physics and Astronomy, University of Birmingham, Birmingham, UK}%
\affiliation{\textsuperscript{2}Materials Engineering Division, Toyota Motor Europe,  Hoge Wei 33, 1930, Zaventum (Belgium)}%
\affiliation{\textsuperscript{3}Ulm University, Ulm, Germany}%
\affiliation{\textsuperscript{4}Institute of Quantum Technologies, German Aerospace Center (DLR), Ulm, Germany}%

\date{\today}
\begin{abstract}
We study the dynamics of a magneto-optical trap (MOT) operating at high-bandwidth. We find the absolute importance of high recapture efficiency between cycles to maintain a practical atom number. We develop a simple model accounting for MOT trapping forces and pressure induced collisions and validate with experimental data using $\mathrm{{}^{87}Rb}$. This is then applied to quantum sensing predicting a shot noise limited sensitivity of $\mathrm{\SI{1E-7}{}\frac{g}{\sqrt{Hz}}}$ for a gravimeter at 100 Hz operation. The results are useful for understanding MOT operation at high-bandwidth, particularly in the context of developing mobile high-bandwidth quantum inertial sensors targeting dynamic environments and navigation applications.
\end{abstract}
\maketitle
\section{INTRODUCTION}
The magneto-optical trap (MOT) has been the workhorse of cold atomic and molecular physics since its first demonstration~\cite{raab1987trapping, chu1985three}. It efficiently cools and traps target species to a sub-millikelvin temperature and is indispensable to the generation of quantum gases, i.e. BEC and degenerate Fermi gas~\cite{anderson1995observation, o2002observation}. The exploration of these fields has resulted in numerous applications in fundamental research and increasingly real-world scenarios such as metrology~\cite{ludlow2015optical}, sensing~\cite{kasevich1991atomic}, quantum simulation~\cite{gross2017quantum, scholl2021quantum}, quantum information processing~\cite{henriet2020quantum, kaufman2021quantum} and so on. Despite the remarkable progress in cold atom physics over the past few decades, most experiments are still conducted in laboratory settings due to the optical, radiofrequency and vacuum requirements for generating and manipulating cold atoms. However, the potential of cold atom technology has been increasingly recognised with efforts made to move experiments out of the laboratory for real-world benefits.


Notably, this trend is evident in the area of quantum gravity sensing, with various demonstrator systems performing trials in different application environments \cite{bongs2019taking, stray2022quantum, bidel2018absolute, menoret2018gravity}. Promising application areas include geophysics, space, civil engineering and oil and mineral prospecting. The potential of the technology is based on its inherent and unparalleled sensitivity, along with the capability of providing drift-free measurements compared to classical approaches. Inertial navigation presents another promising application area for this technology. However, its practical implementation is hindered by the low sampling rate or bandwidth of quantum sensors making them less suited to highly dynamic environments. This limitation primarily arises from the time required for atomic sample preparation, which mainly involves loading the atomic trap, also known as the MOT loading time. As a result, bandwidth is typically limited to roughly 1~Hz. To increase bandwidth, there are various approaches available. One such method is to perform interleaved measurements, starting the next measurement while the previous one is still underway. This approach has demonstrated sampling rates of 3.75 Hz with a measurement time of 801 ms, but it relies on a long drop distance, resulting in a large form factor \cite{savoie2018interleaved}. While sensitive, this implementation competes with the goal of creating small, robust, deployable devices and does not significantly increase bandwidth. Another approach involves using sequential measurements with a considerably reduced cycle time. This method has the potential to increase measurement bandwidth while minimising dead time due to replenishing trapped atoms between cycles. This approach trades bandwidth for reduced sensitivity and system demands. However, achieving 100 Hz operation restricts the cycle time to 10 ms, leaving only a few milliseconds for loading. Consequently, this approach utilises a short drop distance to maintain a high atom number. This smaller displacement ensures that most atoms can be recaptured between cycles, leading to a significant bandwidth increase. Alternatively, one could consider a short loading time with a long measurement time and adopt a 2D MOT or Zeeman slower to enhance the loading rate \cite{joffe1993transverse, dieckmann1998two}. However, this approach will also conflict with the desire for simpler, compact deployable systems.

Quantum sensing is not widely explored at high-bandwidth although some atom interferometry has been performed, achieving sensitivities at the $\sim$µ$\SI{}{g/\sqrt{Hz}}$ level \cite{mcguinness2012high, rakholia2014dual, lee2022compact, adams2021development}. This raises the question of how MOT dynamics and bandwidth are fundamentally connected and the implications for quantum sensing. In this paper, we explore high-bandwidth MOT dynamics in detail, making connections between MOT theory and experimental observations. We build a simple model and validate with experimental data before discussing the critical nature of efficient recapture; optimum parameters and limitations of the mechanism are also explored. The results are then applied to quantum sensing exploring the sensitivity performance limits of a high-bandwidth atom interferometer. This work highlights the utility of simple MOT physics in predicting the feasibility of MOT generation for a given bandwidth, duty cycle, trap size and other cloud properties. Study is performed with the $\mathrm{{}^{87}Rb \, D_{2} \, (5^{2}S_{1/2} \rightarrow 5^{2}P_{3/2})}$ transition. However, general findings apply to a broader range of cold atom experiments targeting higher bandwidth operation.

\section{MODEL} \label{sec:SINGLE ATOM MODEL}
\begin{figure}[H]
  \centering
\includegraphics[width=0.43\textwidth, trim={100 0 100 65},clip]{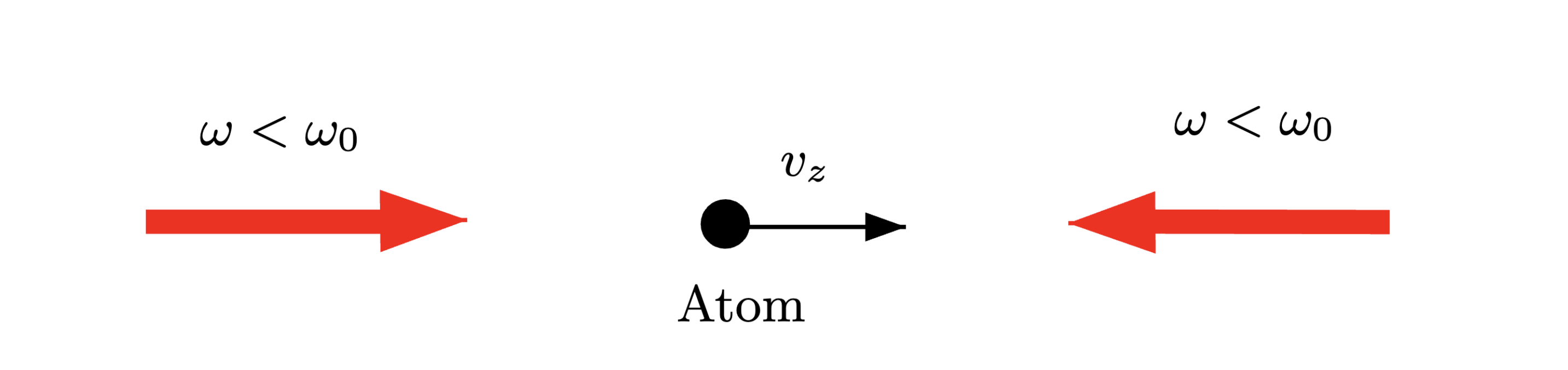}
  \caption{Two counter-propagating laser beams of frequency $\mathrm{\omega}$ less than the atomic resonance frequency $\mathrm{\omega_0}$ illustrating 1D optical molasses. Atom propagates with velocity $v_{z}$ towards the rightmost beam.}
  \label{fig:1D_OM}
\end{figure}

\begin{figure}[H]
  \centering
  \includegraphics[width=0.47\textwidth, trim={5 0 40 65},clip]{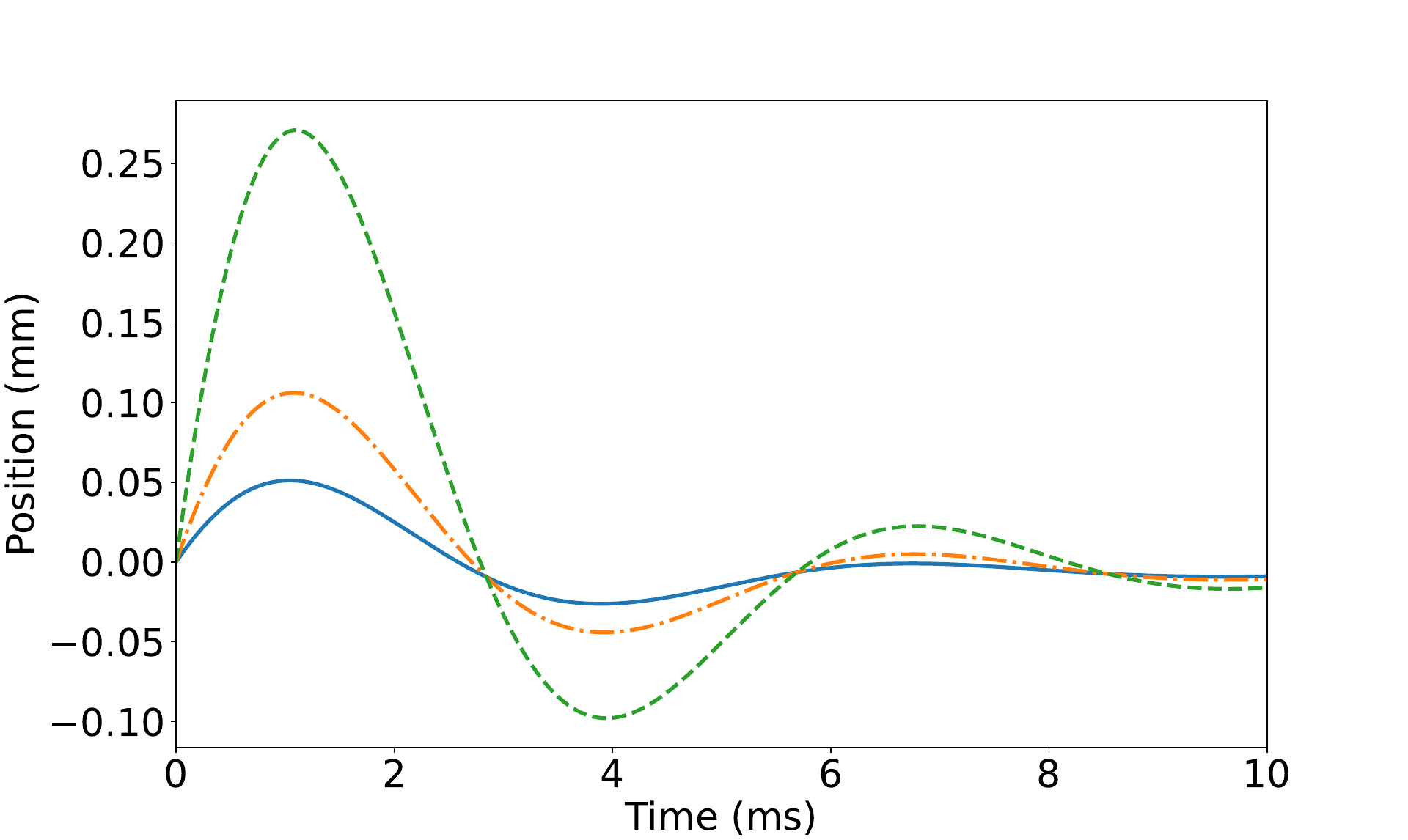}
  \caption{Numerical simulation of single atom trajectories for $\mathrm{{}^{87}Rb}$ atoms with variable initial velocities illustrating under-damped motion occurring for s = 1, $\Delta = -3$, A = 16 G/cm. Initial velocity $\mathrm{v_0}$ ($\SI{}{\metre \second^{-1}}$): 0.5 (green dashed), 0.2 (orange dash-dotted), 0.1 (blue solid).}
  \label{fig:Rb_87_SHM_DEMO}
\end{figure}

To simulate MOT dynamics we adopt the low-intensity theory of optical molasses for a two level atom in 1D illustrated in Fig. \ref{fig:1D_OM} \cite{metcalf1999laser}. This framework can be extended to obtain an expression for the MOT restoring force: $\delta$ corresponds to the detuning from resonance, the $\pm$ subscript accounts for the different detunings of the right and left directed beams, s denotes the saturation parameter and $\Gamma$ is the natural linewidth of the transition. This force is numerically integrated to simulate atomic trajectories. Fig. \ref{fig:Rb_87_SHM_DEMO} demonstrates the MOT restoring force acting on individual $\mathrm{{}^{87}Rb}$ atoms with different initial velocities. This work concerns the $\mathrm{{}^{87}Rb \, D_{2} \, (5^{2}S_{1/2} \rightarrow 5^{2}P_{3/2})}$ transition for which $\mathrm{\Gamma = 2 \pi \times 6.065(9)}$ MHz and $\lambda = 780.241$ nm.

\begin{equation}
        \mathrm{F_{MOT} = \hbar k \frac{\Gamma}{2} \Bigg[\frac{s}{1 + s + (\frac{2\delta_{+}}{\Gamma})^2} -  \frac{s}{1 + s + (\frac{2\delta_{-}}{\Gamma})^2}} \Bigg],
        \label{eq:optical_molasses_full}
\end{equation}

\section{DYNAMICS}
\subsection{INTENSITY DEPENDENCE}\label{sec:Restoration_losses}
For modelling purposes, a simulation cycle is split into two distinct regimes, drop and recapture. For lower bandwidth applications, requirements on MOT loading time are less stringent and so after dropping atoms, loading from background vapour is standard. 
The timescale for this is pressure dependent but typically takes a few 100 ms. Consequently, efficient recapture of atoms between cycles is essential for high-bandwidth operation. The recapture efficiency will not be $100 \%$ but the atom number does not decay to zero as atoms are loaded from the background vapour during recapture. There are two main mechanisms inhibiting recapture; the finite MOT restoring time and collisions between atoms in the MOT and the background vapour.

\begin{figure}[H]
  \centering
  \includegraphics[width=0.47\textwidth, trim={13 5 60 62},clip]{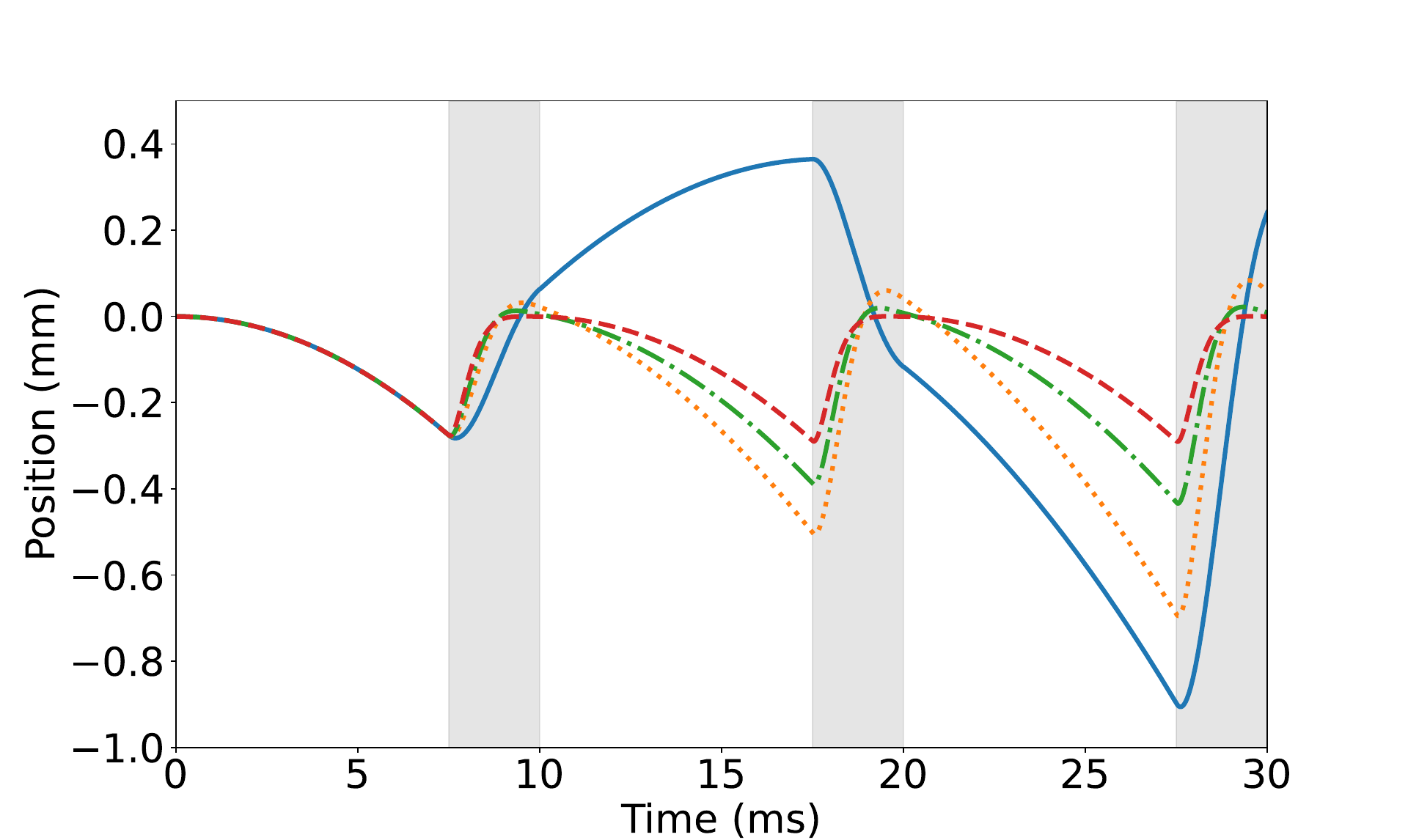}
  \caption{Single atom trajectories in a 100 Hz $\mathrm{{}^{87}Rb}$ MOT for variable intensity. s = 1 (blue solid), 3 (yellow dotted), 5 (green dash-dot) and 10 (red dashed). $\Delta = -3$, duty cycle = 0.75, A = 16 G/cm. The white and grey regions correspond to the drop and recapture phases respectively.}
  \label{fig:Rb_TRAJECTORIES}
\end{figure}

We start by considering the finite restoration time. During freefall atoms move primarily along the vertical and so trajectories are modelled in 1D. For high-bandwidth applications the drop time ($\mathrm{T_{drop}}$) will be $\sim \SI{5}{\milli \second}$ leading to an atom falling $\SI{0.13}{\milli \metre}$. Given a typical trap radius of $\sim \SI{5}{\milli \metre}$, an atom will not fall far from the trap centre. However, despite this short distance, the recapture time is still finite limited by the restoring force towards the MOT centre. Fig. \ref{fig:Rb_TRAJECTORIES} shows a numerical simulation of single atom trajectories over multiple cycles, highlighting that for insufficient power the restoring force is too weak and the atom will not be recaptured. This can be seen in the loss of periodicity for the s = 1 trajectory. Therefore, to maximise bandwidth in experiments, an intensity significantly above the saturation intensity is required to minimise recapture time.

\subsection{TEMPERATURE DEPENDENCE}
To extend this, the dynamics of an atomic cloud are explored by simulating a 1000 atoms with numerical trajectories similar to those in Fig. \ref{fig:Rb_TRAJECTORIES}. The atomic positions and velocities are normally distributed with $\mathrm{\sigma_{MOT}}$ and $\mathrm{\sigma_{v}}$ respectively. $\mathrm{\sigma_{MOT}}$ is the cloud radius and $\mathrm{\sigma_{v} = \sqrt{k_{B}T_{MOT}/m_{atom}}}$ is the cloud's velocity spread where, $\mathrm{T_{MOT}}$ is the cloud temperature and $\mathrm{m_{atom}}$ is the mass of a single atom. To quantify capture, an atom is considered trapped if its final position is $\mathrm{|x_{f}|< 0.1}$ mm from the trap centre and its final speed is $\mathrm{|v_{f}|< \sigma_{v \, Doppler}}$, where $\mathrm{\sigma_{v \, Doppler}}$ is the Doppler velocity.  For cooling on the $\mathrm{{}^{87}}$Rb $\mathrm{D_{2}}$ line, the Doppler cooling limit, $\mathrm{T_{D}}$ = 140 µK, giving $\mathrm{\sigma_{v \, Doppler} = \SI{0.12}{\metre \second^{-1}}}$ \cite{metcalf1999laser}. The fraction of atoms satisfying the capture criteria at the end of the cycle is the restored fraction, $\mathrm{P_{restored}}$. Unless stated, we fix our bandwidth at $\SI{100}{\hertz}$ giving a cycle length of 10 ms. Increasing duty cycle increases the drop time and reduces the recapture time. When the recapture time is $<3$ ms, there is insufficient time to restore atoms to the MOT centre and the recapture efficiency declines. The restored fraction tends to a finite value for short recapture times ($\sim$ 0.05). This results from the spatial extent of the MOT with respect to the capture region. For short recapture times, a fraction of atoms have not left the capture criteria region and are considered recaptured. Furthermore, our simple model applies a Gaussian intensity profile across the 1D trap and so for higher temperatures and longer drop times, atoms move further away from the central most intense region and experience weaker restoring forces. In general, low temperature is critical for cold-atom experiments with our simulations highlighting why this can aid recapture and bandwidth. 

\begin{figure}[H]
  \centering
  \includegraphics[width=0.47\textwidth, trim={22 4 30 5},clip]{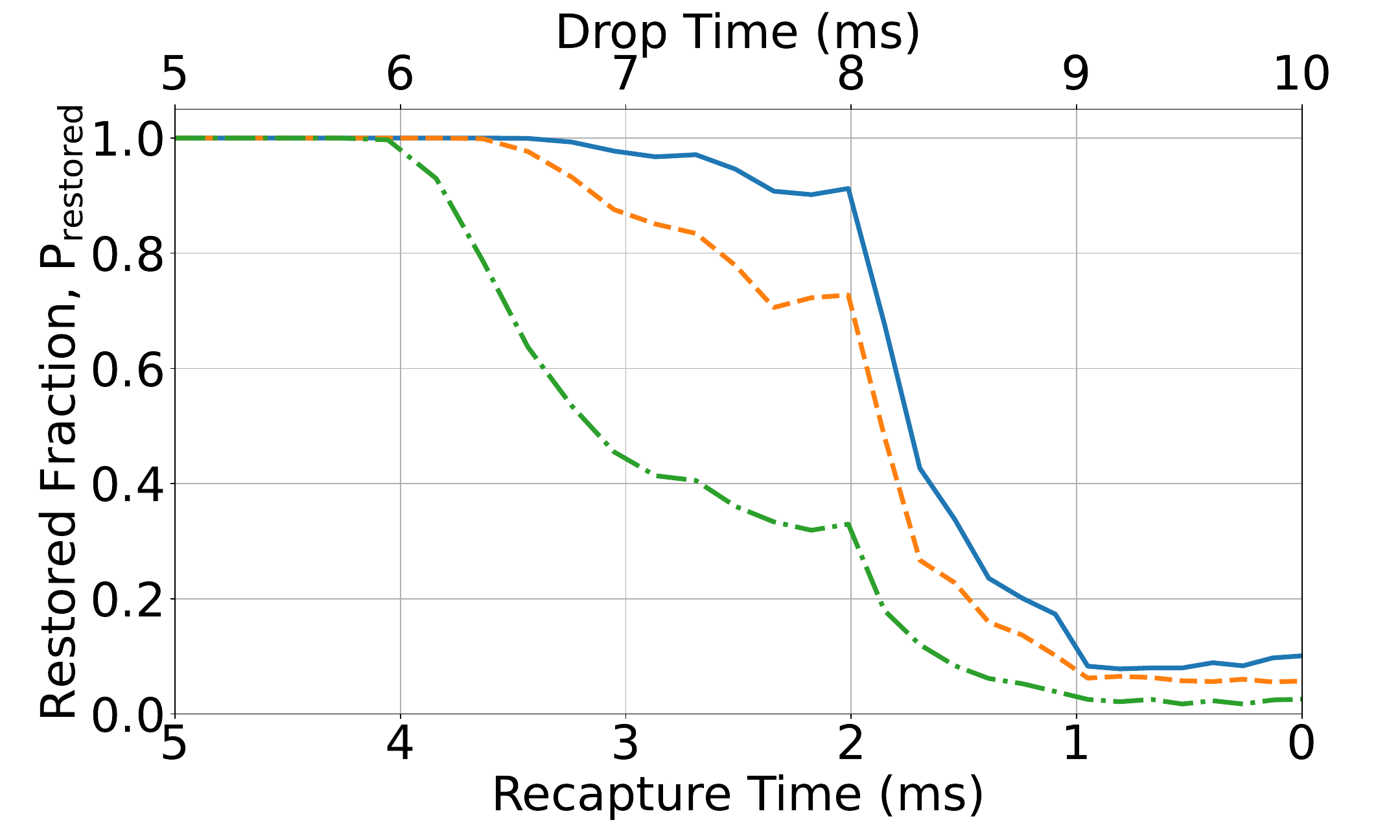}
  \caption{Simulating restored atom fraction for a cloud of $\mathrm{{}^{87}Rb}$ atoms in a 100 Hz MOT for variable duty cycle and cloud temperature. $\mathrm{T_{MOT}}$: 10 µK (blue solid), 100 µK (orange dashed), 1000 µK (green dash-dot).}
  \label{fig:RECAPTURE_FRACTION_TEMPERATURE}
\end{figure}

\begin{figure}[H]
  \centering
  \includegraphics[width=0.47\textwidth,trim={2 2 0
  50},clip]{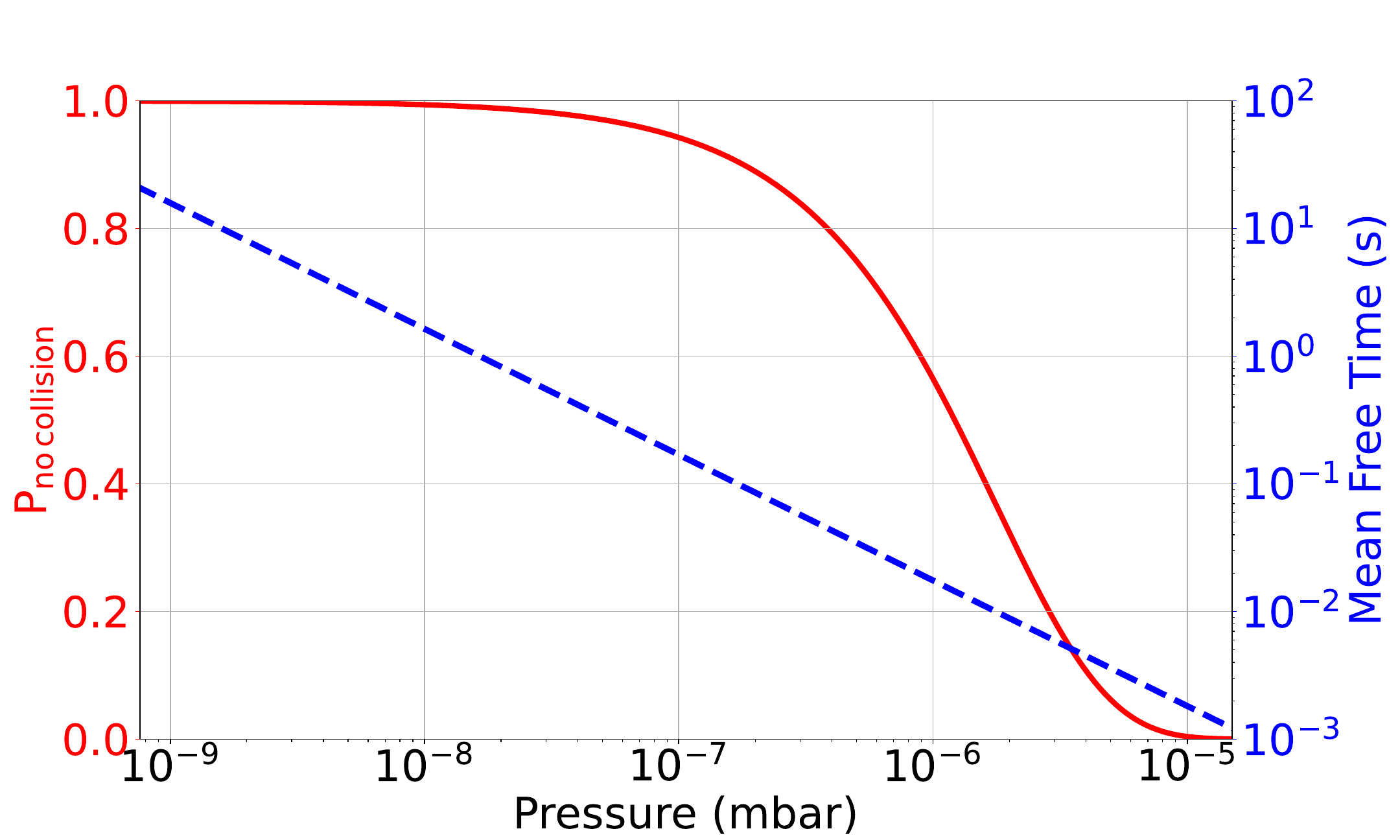}
  \caption{$\mathrm{P_{no\,collision}}$ (red solid) and mean free time (blue dashed) for variable pressure for $\mathrm{T_{cycle} = \SI{10}{\milli \second}}$ \cite{rapol2001loading}.}
  \label{fig:COLLISIONS}
\end{figure}

\subsection{PRESSURE DEPENDENCE}
During an operational cycle, atoms in the cloud can also be lost through collisions with atoms in the background vapour. The probability of this not occurring for an atom during a cycle is given by $\mathrm{P_{no\,collision}}$ in Eq. (\ref{eq:no_BG_col}). $\mathrm{\tau}$ is the mean free collision time and $\mathrm{T_{cycle}}$ is the time for a complete cycle (drop and recapture) as atoms can be lost from background collisions throughout an entire cycle.
\begin{equation}
    \mathrm{P_{no\,collision} = e^{-\frac{T_{cycle}}{\tau}}}.
    \label{eq:no_BG_col}
\end{equation}

For recapture times  $ > \SI{3}{\milli \second}$, restoration losses are typically negligible ($\mathrm{P_{restored}} = 1$) and so Eq. (\ref{eq:no_BG_col}) effectively represents the recaptured atom fraction for a single shot. Unless stated, we use MOT parameters of: $\mathrm{s = 3}$, $\mathrm{\Delta = -3}$, A = 14 G/cm, $\mathrm{T_{MOT}}$ = 300 µK, $\mathrm{\sigma_{MOT}}$ = 0.5 mm, $\mathrm{4\sigma_{r}}$ = 20 mm ($\mathrm{1/e^{2}}$) diameter, Vapour Pressure = $\SI{2.9e-7}{}$ mbar, R = $\SI{4.5e9}{\second^{-1}}$, $\mathrm{L = 16.0 \, s^{-1}}$, $\mathrm{\sigma_{0} = \SI{1e-13}{\centi \metre^2}}$, $\mathrm{C_{v} = \SI{21}{\metre \second^{-1}}}$. $\mathrm{\sigma_{r}}$ defines the trap size, $\mathrm{C_{v}}$ is the capture velocity and R and L define the MOT loading and loss rates respectively. A defines the trap field gradient and $\mathrm{\sigma_{0}}$ defines the collision cross section. More explicit details on these parameters will be given in the subsequent section. Fig. \ref{fig:COLLISIONS} shows the results of computing $\mathrm{P_{no\,collision}}$ and the mean free time over the $10^{-9} - 10^{-6}$ mbar range. For pressures approaching $10^{-6}$ mbar, the collision timescale is comparable to the cycle time, reducing the recaptured fraction significantly. Note, modelling only considers background collisions with $\mathrm{{}^{87}Rb}$ atoms and assumes the absence of other species.

\section{ATOM NUMBER}
\subsection{MOT LOADING}
The rate of change of atoms in the MOT is given by the balance between loading and loss of atoms, integrating this gives the number of atoms after loading for a period of time, t in Eq. (\ref{eq:MOT_LOADING_EQUATION}). 

\begin{figure}[H]
  \centering
  \includegraphics[width=.45\textwidth,trim={2 2 8
  7},clip]{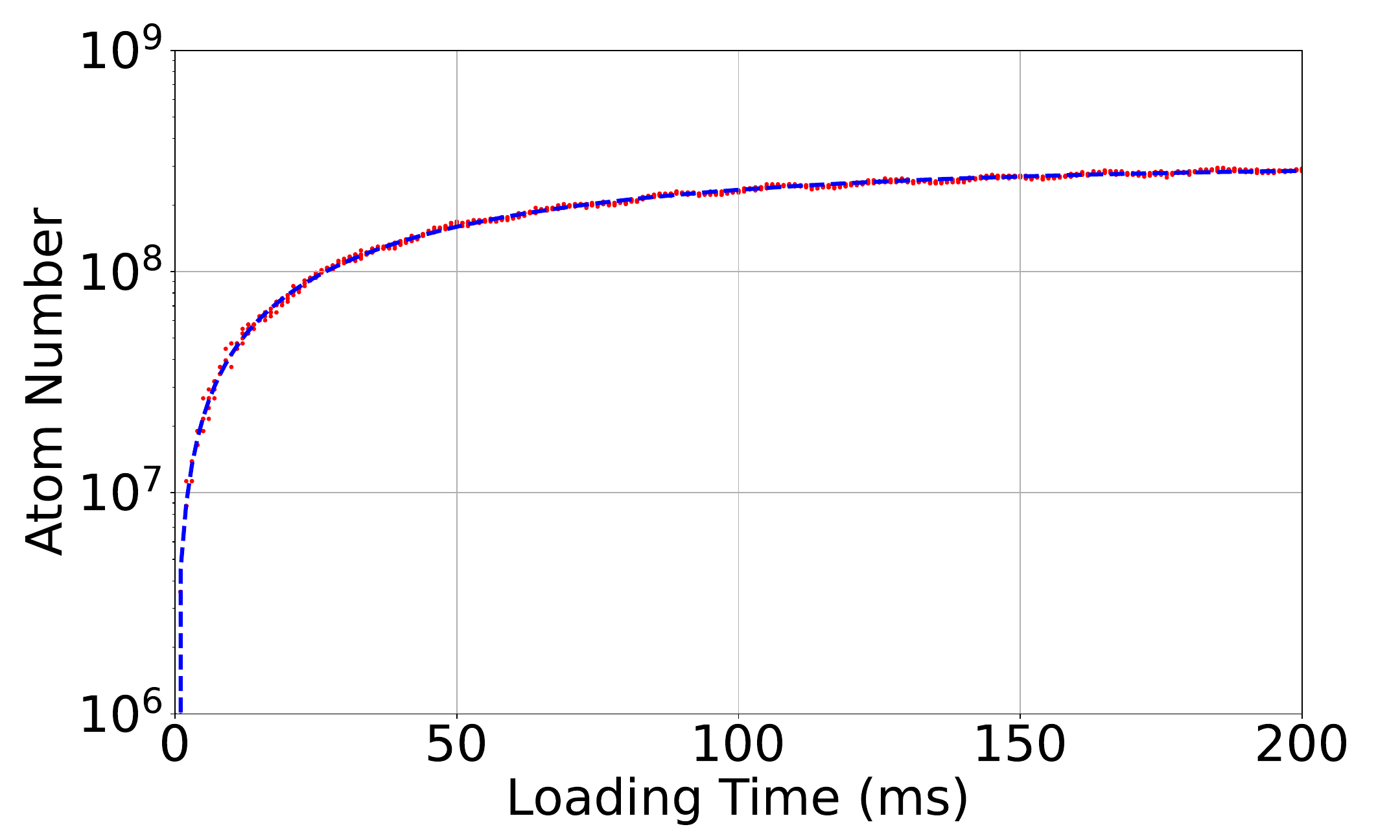}
  \caption{Experimental MOT loading data. The following parameters are extracted, R = $\SI{4.5e9}{\second^{-1}}$, $\mathrm{L = 16.0 \, s^{-1}}$ and a $\mathrm{{}^{87}Rb}$ vapour pressure of $\SI{2.9e-7}{}$ mbar.}
  \label{fig:LOADING_DEMO}
\end{figure}

R and L are the loading and loss rate of the MOT and are given by Eqs. (\ref{eq:LOADING_RATE}) and (\ref{eq:LOSS_RATE}) respectively. $\mathrm{A_{s}}$ is the trap surface area ($\mathrm{4\pi\sigma_{r}^2}$), the capture velocity, $\mathrm{C_{v}}$ is assumed to be $\SI{21}{\metre \second^{-1}}$ - see appendix \ref{sec:Capture Velocity} for details. $\mathrm{n_{b}}$ is the number density of particles in the background vapour, $\mathrm{\sigma_{0}}$ is the collision cross section and $\mathrm{v_{th}}$ is the average thermal velocity of the background gas. The number density of the particles is calculated from the ideal gas equation $\mathrm{n_{b} = \frac{P}{k T}}$ with the vapour pressure obtained from the model in \cite{steck2001rubidium}.

\begin{subequations}
\begin{gather}
    \label{eq:MOT_LOADING_EQUATION} \mathrm{N(t) = \frac{R}{L}(1 - e^{-Lt})}. \\
    \label{eq:LOADING_RATE}
    \mathrm{R = \frac{2 A_s C_{v}^4 n_{b}}{\pi^2 v_{th}^3}}.\\
    \label{eq:LOSS_RATE}
        \mathrm{L = \frac{1}{\tau} = n_{b} \sigma_{0} v_{th}}
\end{gather}
\end{subequations}

The rate equation sometimes includes an additional loss for inelastic collisions between atoms in the MOT. This changes the loss rate to $\mathrm{L \rightarrow L + \beta \bar{n}}$, where $\mathrm{\bar{n}}$ is the mean cloud density and $\beta$ is a constant characterising this mechanism. This implies that two-body collisions can be neglected if $\mathrm{\beta \bar{n} << L}$. $\mathrm{\beta \sim \SI{1e-11}{\centi \metre^3 \second^{-1}}}$ has been reported for a laser detuning of $\mathrm{\delta = -\Gamma}$ and an intensity of $\mathrm{s \approx 10}$, which are fairly typical operating parameters \cite{gensemer1997trap}. Assuming a MOT of around $10^8$ atoms with a radius of $\mathrm{\SI{1}{\milli \metre}}$ gives a number density of $\mathrm{\bar{n} \sim \SI{1e10}{\centi \metre^{-3}}}$. For typical pressure $\mathrm{L \sim 1-10 \, s^{-1}}$ which is 1-2 orders higher than the two body loss term. This justifies why this term can be neglected in our simulations. For 100 Hz operation the MOT loading time is only a few ms. Even for relatively high pressures in the low $10^{-7}$ mbar range the loading rate is a few $10^{9}$/ms. This means at most $\sim 10^{7}$ atoms can be loaded from the background vapour after a few $\SI{}{\milli \second}$; a small fraction of the steady state population reached in the experimental data in Fig. \ref{fig:LOADING_DEMO}. This highlights how efficient recapture of atoms between cycles is essential for high-bandwidth operation. In this regime MOT composition is recapture dominated with a small contribution from background loading. Consider a high-bandwidth MOT containing $10^{7}$ atoms with a recapture period of $\sim \SI{1}{\milli \second}$. Assuming recapture is $90\%$ efficient with a MOT loading rate of $\mathrm{R \sim 10^{9} \, s^{-1}}$ the atom number will remain steady. By considering losses from the finite restoration time and collisions independently, an iterative equation is formed describing the shot to shot atom number.

\begin{equation}
        \mathrm{N_{i+1}} = \mathrm{N_{i}P_{no\,collision}P_{restored} + \frac{R}{L}(1 - e^{-LT_{Reload}})}.
\label{eq:recaptured_eq}
\end{equation}    

$\mathrm{N_{i}}$ denotes the atom number in the $\mathrm{i^{th}}$ cycle. The first term describes the contribution from recaptured atoms with $\mathrm{P_{no\,collision}P_{restored}}$ representing the constant shot to shot recapture fraction. The second term describes background loading and is the MOT loading equation with terms as defined in Eq. (\ref{eq:MOT_LOADING_EQUATION}). The time for loading and recapture is given by $\mathrm{T_{reload}}$. Iterating until $\mathrm{N_{i + 1}} = \mathrm{N_{i}}$ gives the operational steady state atom number for the MOT. For higher pressure the loading rate is larger and so more atoms are loaded from the background but fewer atoms are recaptured due to more background collisions and vice versa for lower pressure. Steady state corresponds to the point at which the number of atoms lost due to inefficient recapture is perfectly balanced by the atoms loaded from the background vapour. 

\begin{figure}[H]
  \centering
  \includegraphics[width=.44\textwidth,trim={7 0 8
  42},clip]{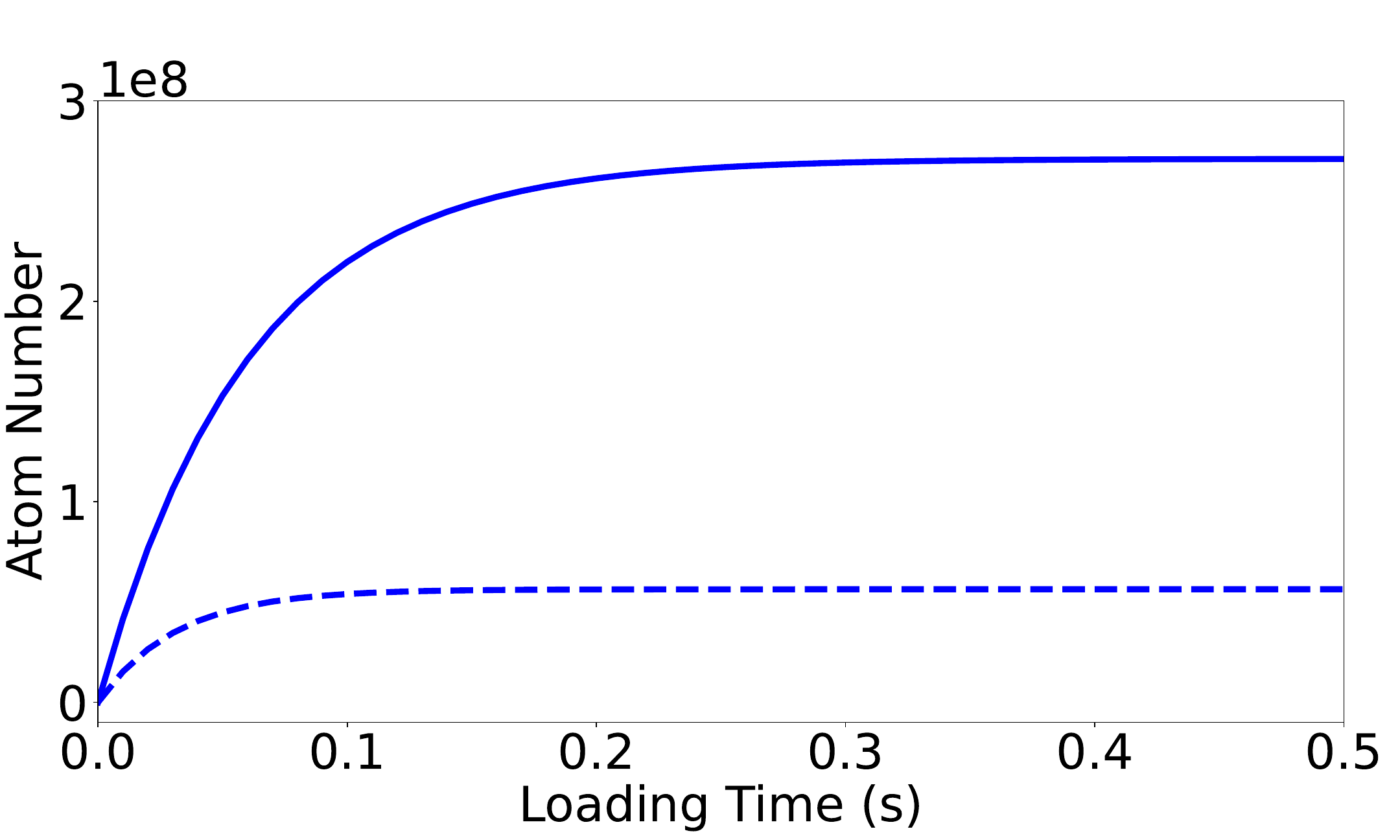}
  \caption{Traditional non-dynamic MOT loading (solid), 100 Hz high-bandwidth MOT loading simulation at a duty cycle of 0.65 (dashed).}
  \label{fig:LOADING_ITERATION}
\end{figure}

In Fig. \ref{fig:LOADING_ITERATION} the behaviour of a traditional MOT is simulated and contrasted with a high-bandwidth MOT with a duty cycle of 0.65. In this configuration there are about $20\%$ the number of atoms when compared with a MOT fully loaded from background vapour. Even with our relatively high pressure, without recapture it would take 10x longer to load this many atoms. This limits bandwidth to at most 30 Hz showing the importance of recapture in maximising bandwidth.

\subsection{DUTY CYCLE}
\begin{figure}[H]
  \centering
  \includegraphics[width=0.45\textwidth,trim={2 1 50
  20},clip]{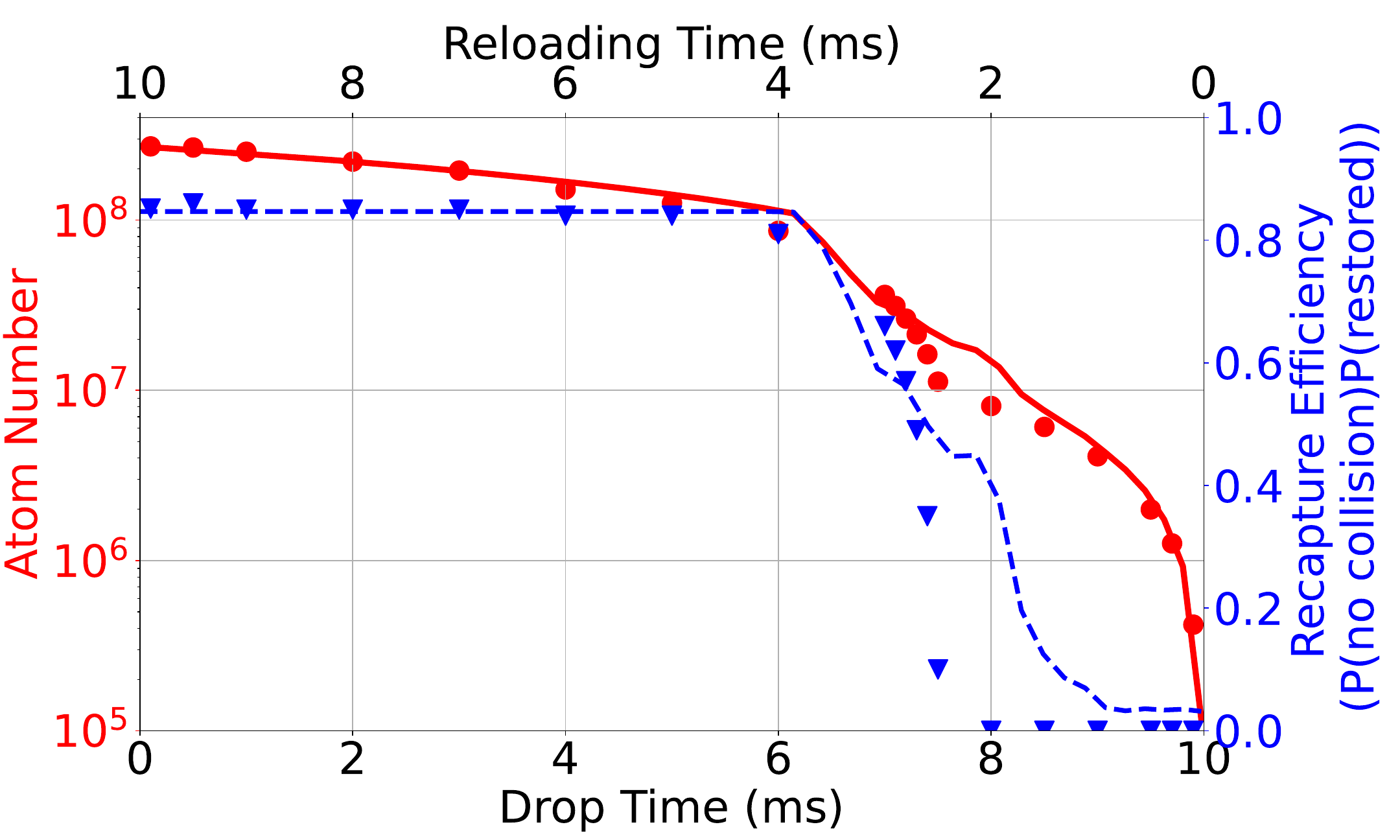}
  \caption{Steady state atom number (red solid) and recapture efficiency $\mathrm{P_{no \, collision}P_{restored}}$ (blue dashed) for a 100 Hz MOT for variable duty cycle. Experimental data points are scattered.}
  \label{fig:100Hz_ATOM_NUMBER}
\end{figure}

A key parameter determining MOT operation is the duty cycle describing the useful fraction of the experimental cycle. In this context it denotes the free-fall time. The remaining portion constitutes time for recapturing and loading atoms back into the trap for the next cycle. Optimising duty cycle is important for experimental applications as increasing measurement time will compromise time available for reloading atoms into the MOT. Naturally, some balance must be achieved within a cycle. To investigate this we vary the parameter experimentally and compare with our simple dynamics model. Fig. \ref{fig:100Hz_ATOM_NUMBER} presents data at 100 Hz bandwidth, as drop time tends to 0 ms the atom number tends towards the value in Fig. \ref{fig:LOADING_ITERATION} for non-dynamic MOT operation. For increasing drop times up to 6 ms the atom number decays gradually as less cycle time is devoted to reloading. In this regime, the recapture efficiency stays constant as the restoration force is sufficient to recapture atoms for reloading time $>$ 3.5 ms ($\mathrm{P_{restored} = 1}$). The imperfect recapture efficiency comes from the pressure induced collisions with the background vapour, $\mathrm{P_{no \, collision}}$ = $85\%$ at 100 Hz. For drop times $> 6.5$ ms the recapture mechanism fails and the atom number declines dramatically with a good fit between model and experimental data. This fit is slightly poorer at 50 Hz but still quite reasonable. Given the 1D model used, further discrepancies might be connected to the 3D nature of the light field, magnetic field and polarisation profiles. To validate our collision model we perform duty cycle scans with fixed cycle times of 2.5, 5, 10 and 20 ms. Using this data we extract the $\mathrm{P_{no \, collision}}$ value as drop time tends to 0 ms and plot against Eq. (\ref{eq:no_BG_col}) for our operating pressure of $\mathrm{\SI{2.9e-7}{}}$ mbar.

\begin{figure}[H]
  \centering
  \includegraphics[width=0.49\textwidth,trim={40 1 45
  20},clip]{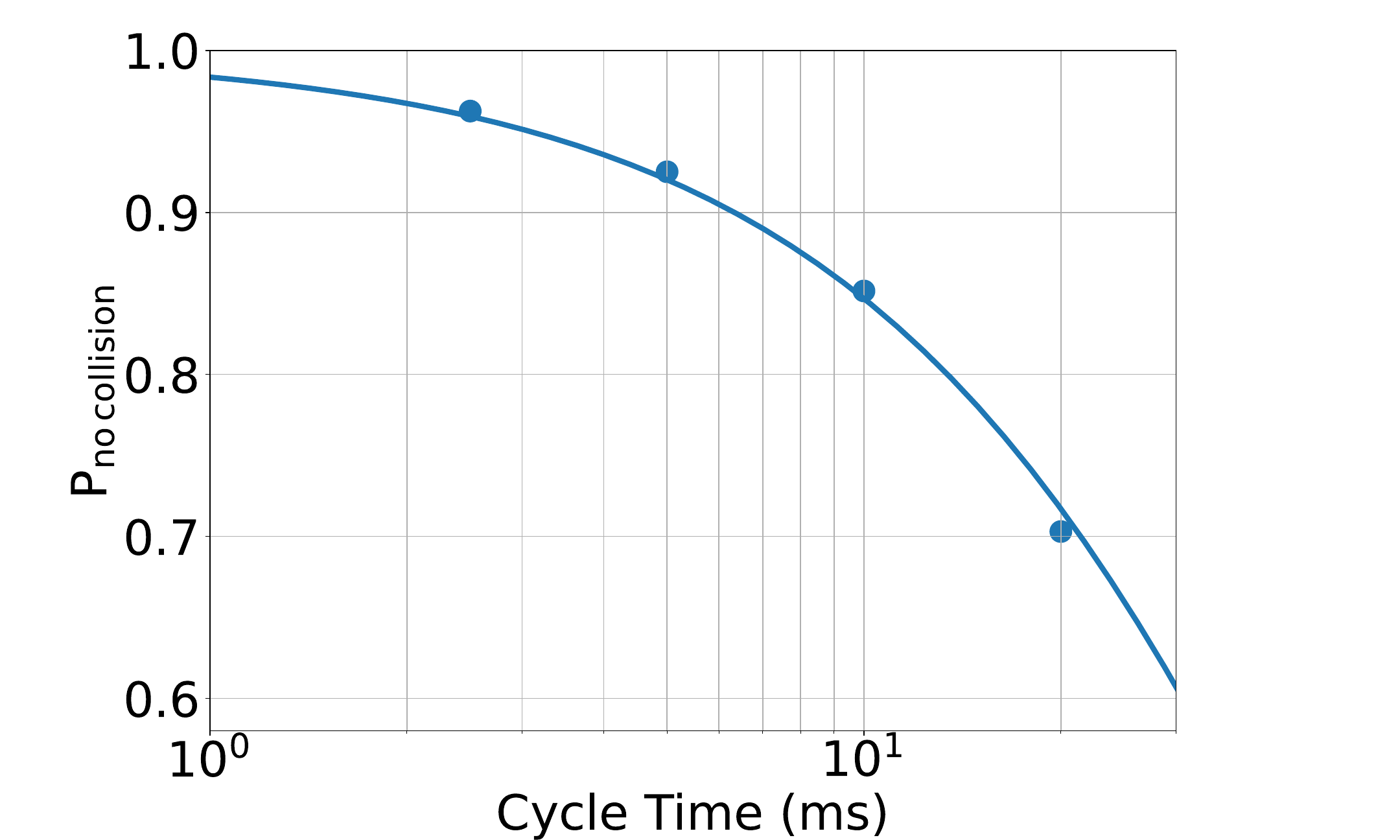}
  \caption{Pressure induced collision model, theoretical model (line), experimental data (points).}
  \label{fig:PRESSURE_MODEL_VALIDATE}
\end{figure}

\begin{figure}[H]
  \centering
  \includegraphics[width=.495\textwidth,trim={47 10 -70
  65},clip]{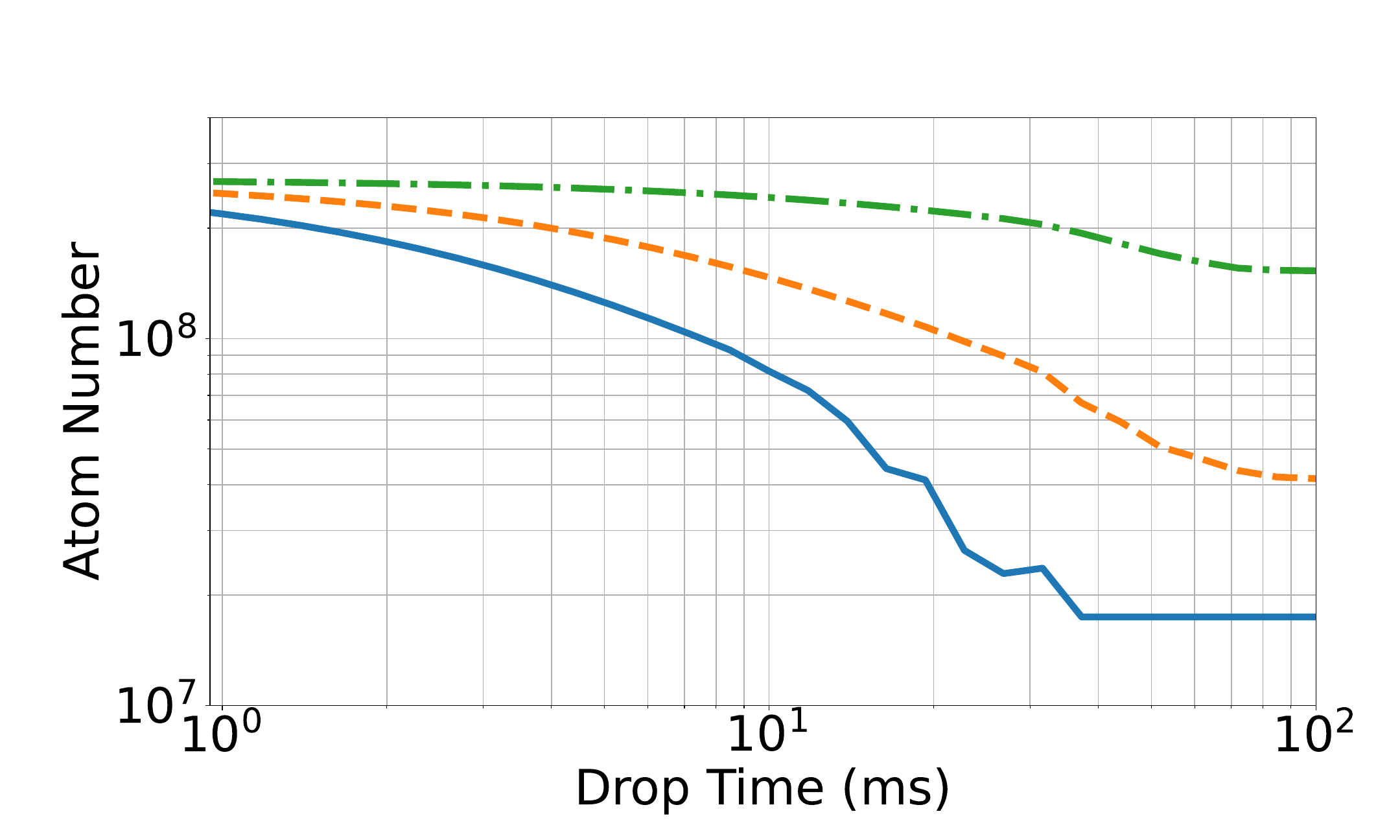}
  \caption{Steady state atom number for variable drop time with a fixed loading time: 4.0 ms (blue solid), 10 ms (orange dashed) 50 ms (green dash-dot).}
  \label{fig:VARYING_DROP_TIME_ATOM_NUMBER}
\end{figure}

Fig. \ref{fig:PRESSURE_MODEL_VALIDATE} presents this data showing a strong fit validating our collision model. To further highlight the importance of recapture we simulate longer drop times with a short reloading time. To model this, the reloading time is fixed, the drop time is incremented and the steady state atom number is computed. After falling $\mathrm{2\mathrm{\sigma_{r}}}$ = 10 mm, an atom will fall out of the trap centre in $\sim \SI{45}{\milli \second}$ as reflected in the decline in Fig. \ref{fig:VARYING_DROP_TIME_ATOM_NUMBER}. For drop times $\ll \SI{45}{\milli \second}$ the dynamics are recapture dominated as atoms do not fall out of the trapping region. For drop times $ > \SI{45}{\milli \second}$ the MOT is no longer in the trapping region and so recapture is not viable. Consequently, the MOT consists entirely of atoms loaded from the background vapour. For longer loading times the drop off is less pronounced highlighting the need for a significant increase in reloading time when leaving the recapture dominated regime. Our model is further validated by calculating and measuring the reloading time for a steady state MOT of $10^{8}$ atoms. As anticipated, the recapture efficiency experiences a decline to zero at 45 ms of drop time. For small drop times the loading time required tends to the MOT restoration time for a $\mathrm{{}^{87}Rb}$ atom ($\sim$ 3 ms) in this regime. When recapture fails, the time required is determined entirely by background loading and is given by $ \frac{\SI{1e8}{}}{\SI{4.5e9}{\second^{-1}}} \sim$ 25 ms. For lower pressures ($\sim 10^{-8}$ mbar) this time will be significantly longer due to the reduced loading rate. Overall, a good fit is observed between the model and experiment. For experiments care is required to ensure sufficient loading time such that recapture is not compromised. Equally, excess time should be avoided to promote measurement bandwidth. To optimise this in different systems analysis similar to Fig. \ref{fig:100Hz_ATOM_NUMBER} could be performed by increasing the duty cycle until a sharp drop off in atomic signal is observed. This reflects the point at which the recapture mechanism fails determining the necessary trap loading time.

\begin{figure}[H]
  \centering
  \includegraphics[width=0.45\textwidth,trim={2 1 50
  5},clip]{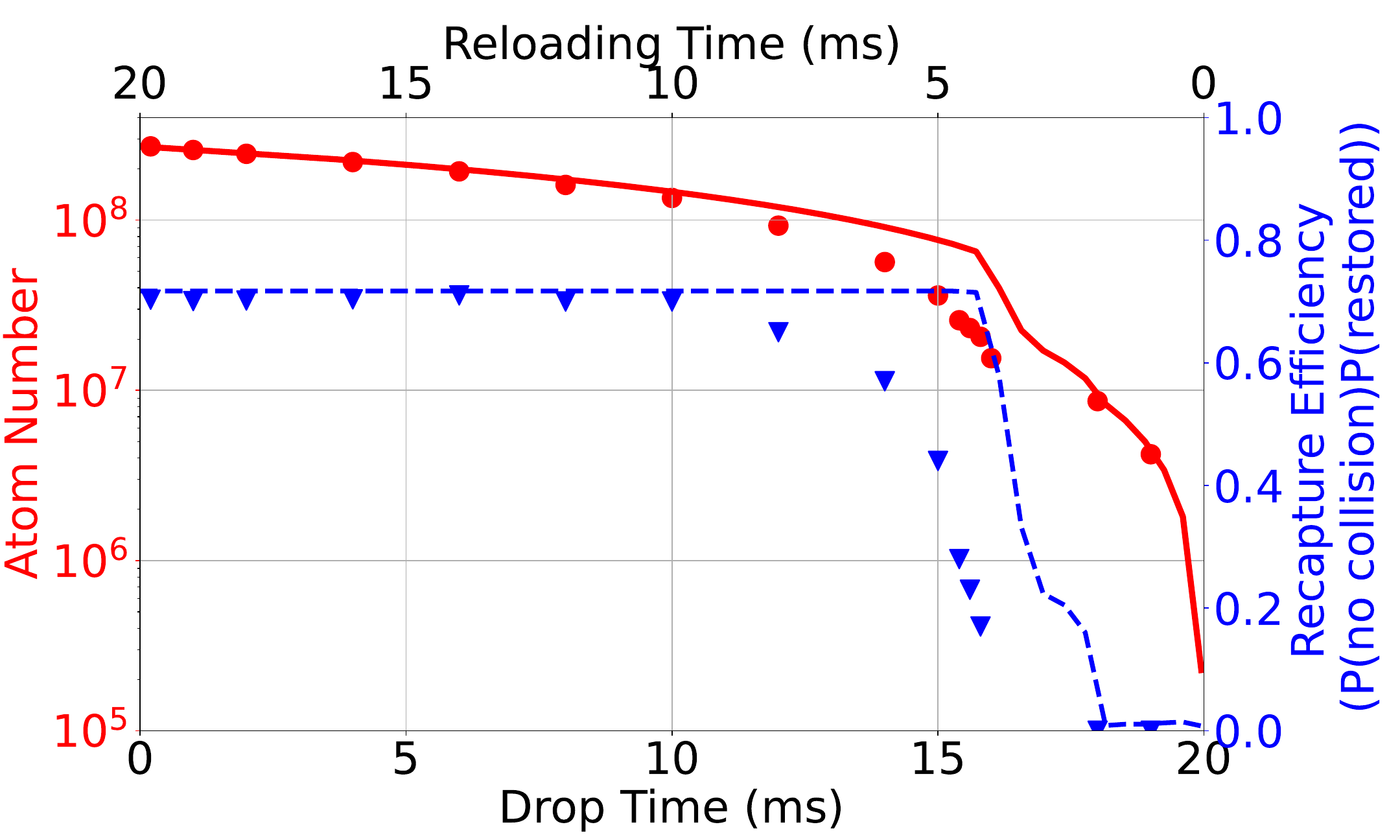}
  \caption{Steady state atom number (red solid) and recapture efficiency $\mathrm{P_{no \, collision}P_{restored}}$ (blue dashed) for a 50 Hz MOT for variable duty cycle. Experimental data points are scattered.}
  \label{fig:50Hz_ATOM_NUMBER}
\end{figure}

\begin{figure}[H]
  \centering
  \includegraphics[width=.475\textwidth,trim={51 2 71
  12},clip]{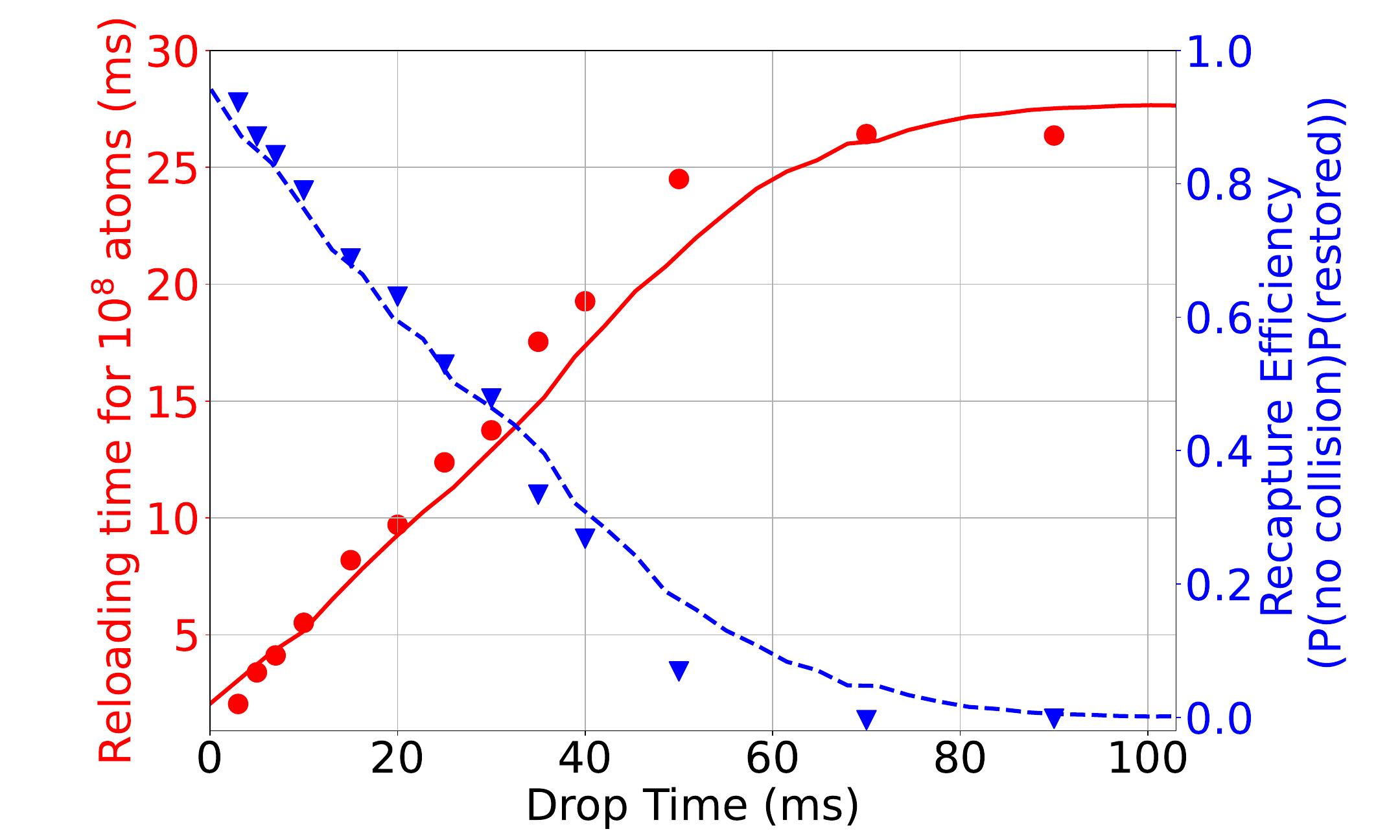}
  \caption{Time to load $10^{8}$ atoms for variable drop time (red solid), recapture efficiency $\mathrm{P_{no \, collision}P_{restored}}$ (blue dashed). Experimental data points are scattered.}
  \label{fig:REQUIRED_RELOADING_1E8}
\end{figure}

\section{DISCUSSION}
\subsection{Application to Quantum Sensing}
\begin{figure}[H]
  \centering
\includegraphics[width=0.475\textwidth, trim={15 5 10 16},clip]{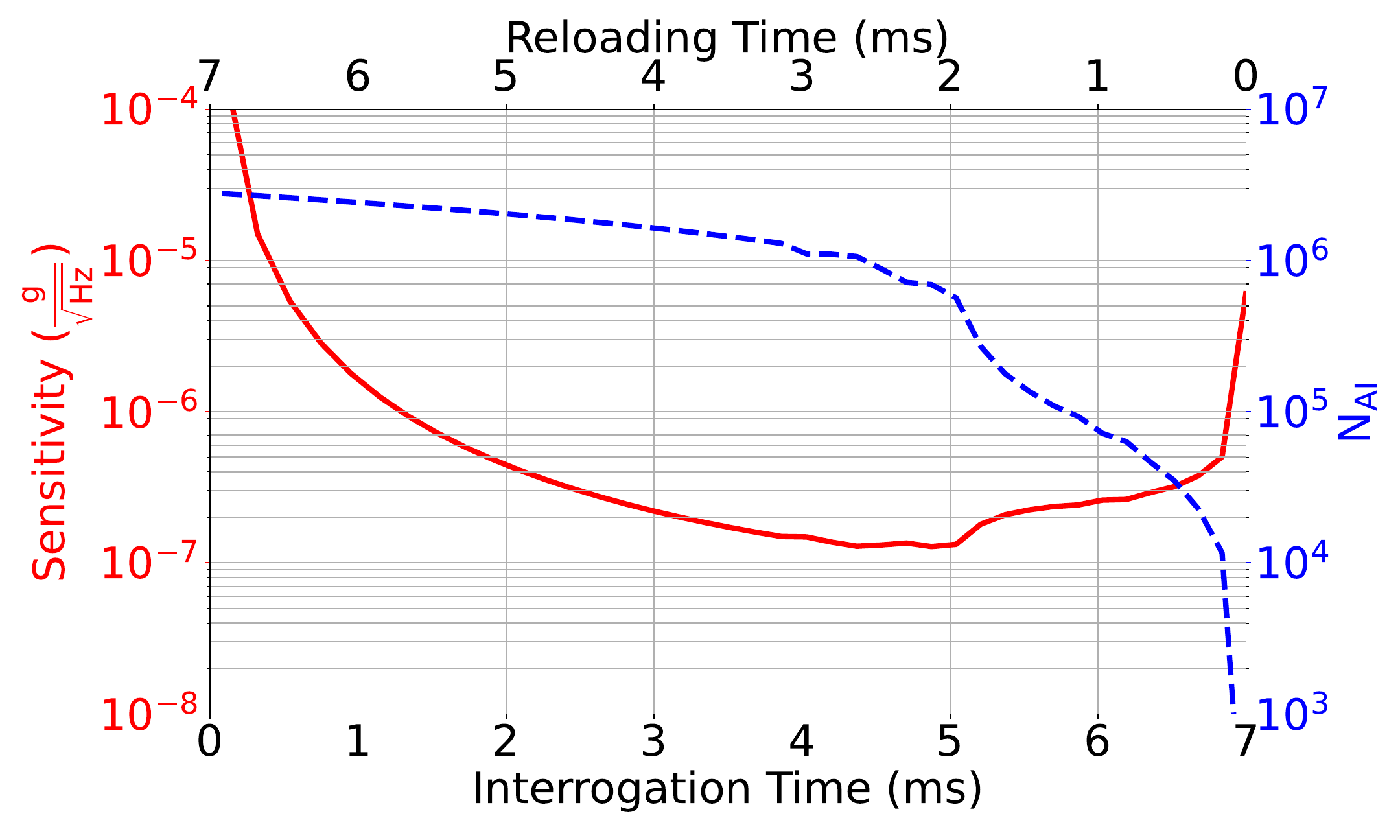}
  \caption{Optimising sensitivity by optimising balance between recapture and interrogation time, sensitivity (red solid), participating atoms (blue dashed). The optimised cycle consists of a 5 ms interrogation, 2 ms recapture and a set 3 ms of additional preparation (cooling, state preparation, launching). AI parameters: F = 100~Hz, $\eta$ = 1, $\mathrm{N_{AI} = 0.01\, N_{MOT}}$.}
  \label{fig:SENSITIVITY_100Hz}
\end{figure}

Having validated our simple model for the high-bandwidth MOT we will now apply this to optimise an application. Atom interferometry (AI) was developed in the early 1990s and offers exceptional sensitivity to rotations and accelerations \cite{carnal1991young}. The technique underpins quantum sensing which shows huge promise for applications in inertial navigation \cite{wright2022cold, adams2021development}. To explore this we predict the sensitivity performance limit of an atom interferometer operating at 100 Hz. Sensitivity is given by $\mathrm{\frac{\delta \phi}{\phi}}$, where $\mathrm{\delta \phi}$ denotes phase noise and $\phi$ is the phase signal accumulated over the interrogation period. The noise on a single measurement $\delta \phi_{s}$ is limited by quantum projection noise $\mathrm{N_{Q} = \sqrt{N_{AI}}}$ and $\mathrm{\delta \phi_{s} = \eta \delta \phi_{Q} = \eta \frac{N_{Q}}{N_{AI}} = \frac{\eta}{\sqrt{N_{AI}}}}$,  where $\mathrm{N_{AI}}$ denotes the number of atoms participating in the interferometer with $\mathrm{\eta \geq 1}$ accounting for excessive detection noise. The operating bandwidth is given by $\mathrm{F = \frac{1}{(T_{i} + T_{p})}}$ where $\mathrm{T_{i} = T_{drop}}$ is the interrogation (drop) time and $\mathrm{T_{p}}$ is the sensor preparation time incorporating reloading, cooling and detection. Using these definitions sensitivity can be expressed as in Eq. (\ref{eq:Sensitivity}). 
\begin{equation}
    \mathrm{S = \frac{4\eta}{k_{e}g\sqrt{N_{AI}}\sqrt{F}T_{i}^2} \approx \SI{2.5e-8}{}\frac{\eta}{\sqrt{N_{AI}}}\frac{\sqrt{F^3}}{(1 - FT_{p})^2}}.
    \label{eq:Sensitivity}
\end{equation}
For optimal sensitivity the duty cycle requires optimisation to balance the recapture and interrogation periods. Assuming a certain bandwidth, duty cycle and shot noise limited detection the only unknown in Eq. (\ref{eq:Sensitivity}) is atoms participating in the interferometer, n. To acquire this the recapture simulation is run for the chosen duty cycle and MOT parameters to obtain the recapture efficiency. The atom number is then computed using Eq. (\ref{eq:recaptured_eq}). A conservative $1\%$ of atoms are assumed to complete the interferometer, $\mathrm{N_{AI} = 0.01 \, N_{MOT}}$. To account for sub-Doppler cooling, state preparation and launching, a 3 ms preparation time is allocated within the cycle time. We also adopt a cloud temperature of 10 µK following sub-Doppler cooling. Fig. \ref{fig:SENSITIVITY_100Hz} shows the sensitivity simulation at 100 Hz operation for variable duty cycle. For lower duty cycles there are more atoms but the sensitivity improvement from increased interrogation time dominates over the reduced atoms. For reloading times $<$ 2 ms the capture processes are inhibited and the atom number falls to zero diminishing sensitivity. Fig. \ref{fig:SENSITIVITY_100Hz} suggests a performance limit of $\SI{1e-7}{\frac{g}{\sqrt{Hz}}}$ at 100 Hz operation. Given the finite recapture time it is interesting to consider optimal sensitivity for variable bandwidth. To explore this the simulation in Fig. \ref{fig:REQUIRED_RELOADING_1E8} is reprocessed. By adding the drop and reloading time together and including an additional 3 ms of preparation time a certain cycle time and therefore bandwidth is defined. For this bandwidth $10^{8}$ atoms are generated and so sensitivity can be computed with Eq. (\ref{eq:Sensitivity}).

\begin{figure}[H]
  \centering \includegraphics[width=0.475\textwidth, trim={8 6 10 2},clip]{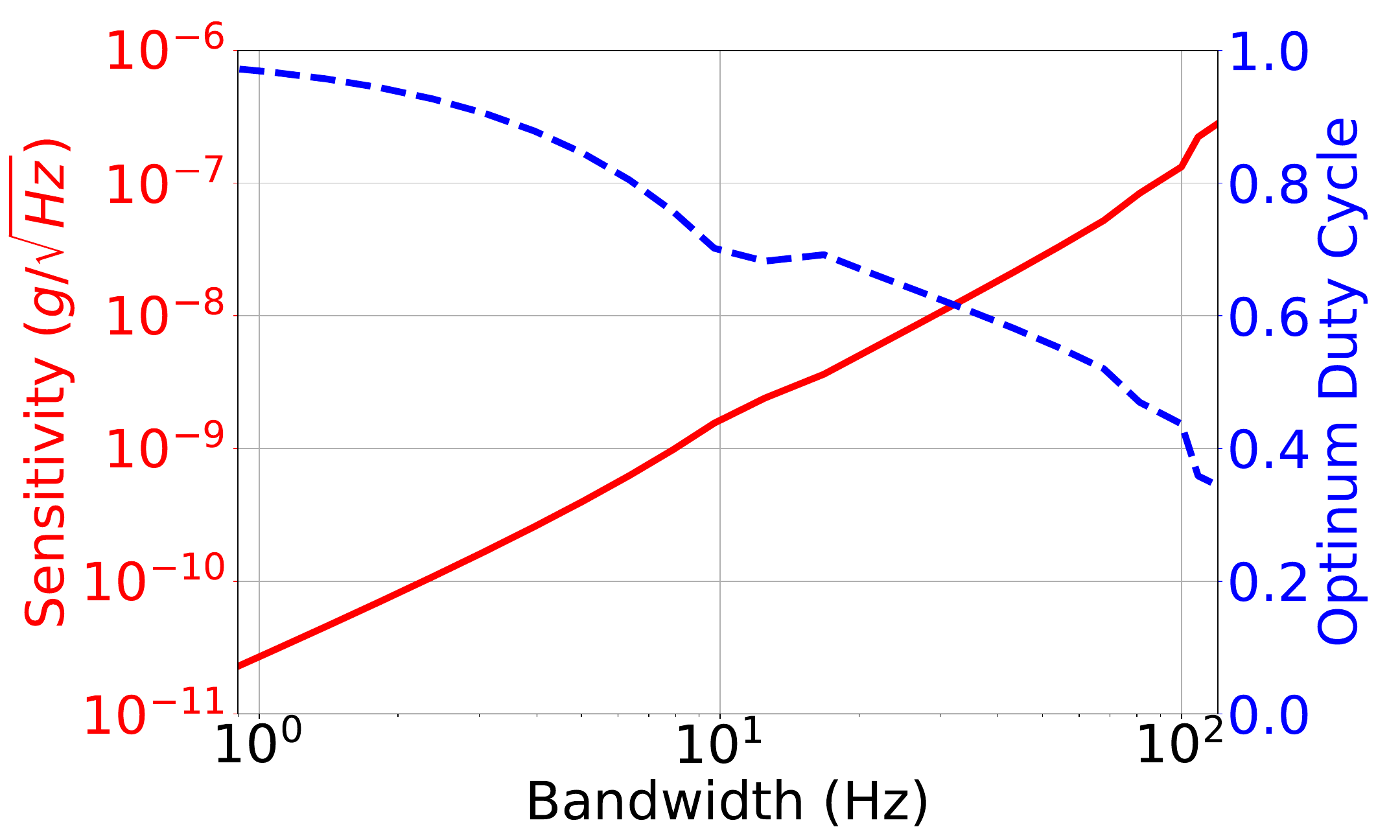}
  \caption{Sensitivity projection for variable bandwidth based on simulation in Fig. \ref{fig:REQUIRED_RELOADING_1E8}. For each bandwidth the cycle consists of an additional 3 ms of preparation.}
  \label{fig:SENSITIVITY_VARIABLE_BW}
\end{figure}

For increasing bandwidth the optimal duty cycle decreases gradually as the necessary reloading time represents a larger fraction of the cycle, see Fig. \ref{fig:SENSITIVITY_VARIABLE_BW}. At a certain bandwidth the cycle time is insufficient to interrogate, recapture and prepare atoms. For short drop time around 2 ms is required to recapture atoms and so with an additional preparation time of $\SI{3}{\milli \second}$ the limiting bandwidth is $\frac{1}{5\SI{}{\milli \second}} \simeq \SI{200}{\hertz}$. Given the performance limits it is worth summarising the advantages, disadvantages and future prospects of the high-bandwidth approach for quantum sensing. Quantum sensors offer low bias and high-stability enabling long term inertial navigation measurements not currently feasibly with classical sensors. High-bandwidth quantum sensors would therefore be attractive for navigation where measurement rates $>$ 100 Hz are needed for operation on mobile platforms. 

As highlighted bandwidth and sensitivity present a compromise although the reduced free-falling distance at high-bandwidth makes the approach compelling for miniaturisation developing devices more robust to challenging environments \cite{lee2022compact}. The $\sim$µ$\SI{}{g/\sqrt{Hz}}$ sensitivity offered at high-bandwidth would be useful for inertial navigation with techniques such as large-momentum transfer potentially offering a route to clawing back sacrificed sensitivity \cite{wilkason2022atom}. Even presently ship-borne measurements have demonstrated sensitivities at the $\sim$µg level \cite{bidel2018absolute}. Moreover, hybrid methods have been implemented to increase bandwidth using a quantum sensor to correct a classical device \cite{cheiney2018navigation}. Further developments could offer potential for absolute positioning on a metre scale independent of environment without satellite navigation. Moreover, high-bandwidth operation would also be desirable for faster civil engineering surveys providing feedback on the condition of water pipes and identifying voids and mine shafts.

\section{CONCLUSIONS}
We show that a simple model simulating atomic trajectories and loss mechanisms performs rather well in explaining experimental MOT dynamics across a range of bandwidths. Traditionally bandwidth is not a primary concern and so traps are loaded to capacity with no concern for recapturing atoms limiting bandwidths to around 1 Hz. In this work we explore the full bandwidth range. At low bandwidth recapture efficiency tends to 0 due to background collisions and atoms falling outside of the trapping region. At high-bandwidth the finite MOT restoring force is critical limiting the recapture time to a few ms for $\mathrm{{}^{87}Rb}$ and imposing a maximum bandwidth for MOT generation. We observe that the model provides a good fit to experimental data across a range of bandwidths accounting for pressure, temperature and spatial considerations of the trap. The model is then applied to quantum sensing projecting a performance limit of $\SI{1e-7}{g/\sqrt{Hz}}$ at 100 Hz. This is computed by optimising duty cycle for a given bandwidth. Based on this it is deemed beneficial to devote cycle time to interrogation provided recapture is not compromised significantly. In summary, this work shows the power of a simple MOT physics model in predicting the feasibility of MOT generation for a given bandwidth, duty cycle and other trap and cloud properties. More generally, the ubiquitous nature of the MOT means this work could be applied to a broad range of experiments using different atomic species particularly for those targeting higher bandwidth operation.

\begin{acknowledgments}
We thank the support of the UK National Quantum Technologies Programme (NQTP) (EP/T001046/1), Defense Science and Technology Laboratory (Dstl) (DSTLXR1000141929) and Toyota Motor Europe. 
\end{acknowledgments}

\bibliography{apssamp}
\appendix
\section{Capture Velocity} \label{sec:Capture Velocity}
\begin{figure}[H]
  \centering
  \includegraphics[width=0.45\textwidth, trim={10 2 5 61},clip]{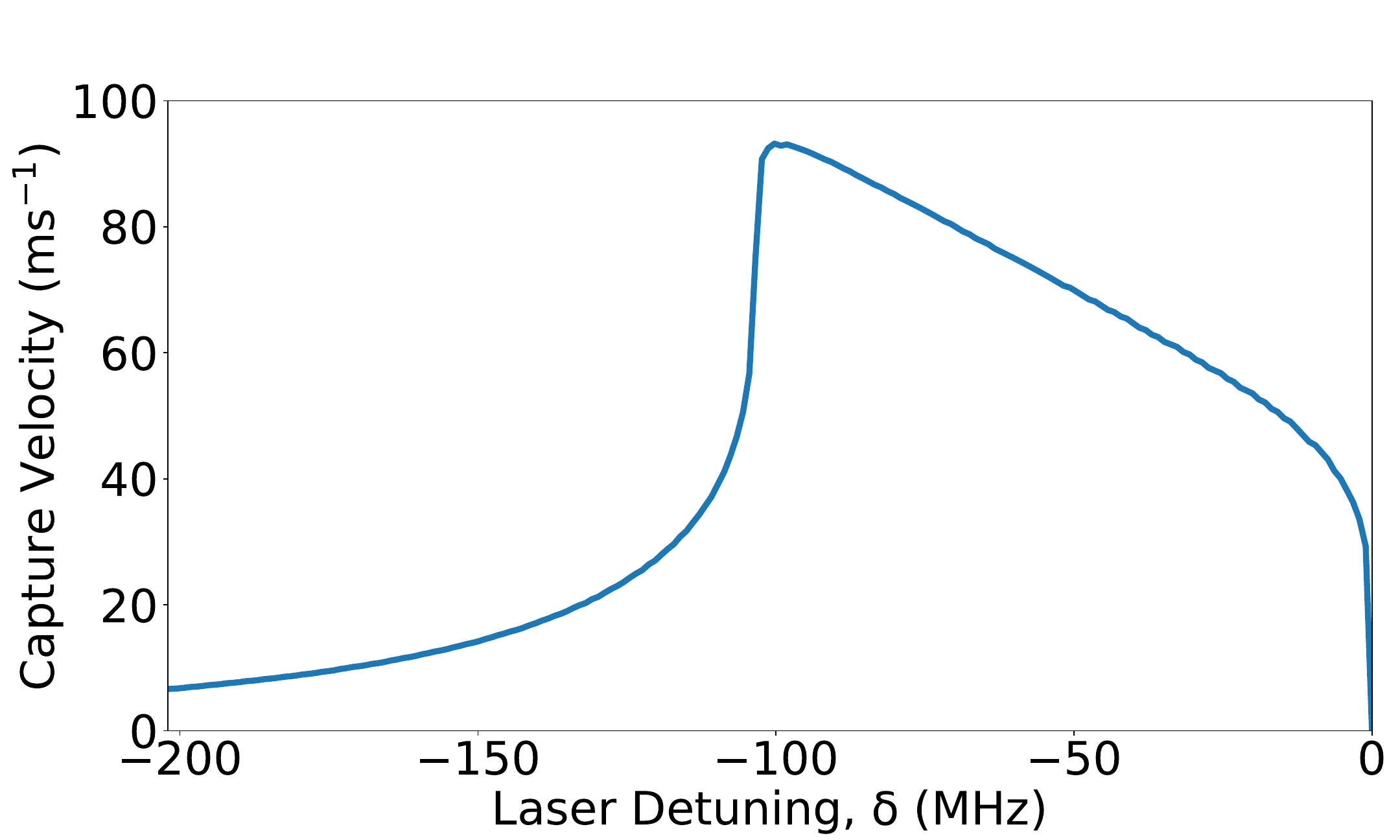}
\caption{Dependence of capture velocity, $\mathrm{C_{v}}$ on detuning for $\mathrm{{}^{23}Na}$. The largest $\mathrm{C_{v}}$ is obtained for a detuning of $\approx -\SI{100}{\mega \hertz} \approx -10 \Gamma$. Simulation time = 100 ms.}
  \label{fig:Na_23_CAPTURE_VELOCITY}
\end{figure}

\begin{figure}[H]
  \centering
  \includegraphics[width=0.46\textwidth, trim={0 3 2 40},clip]{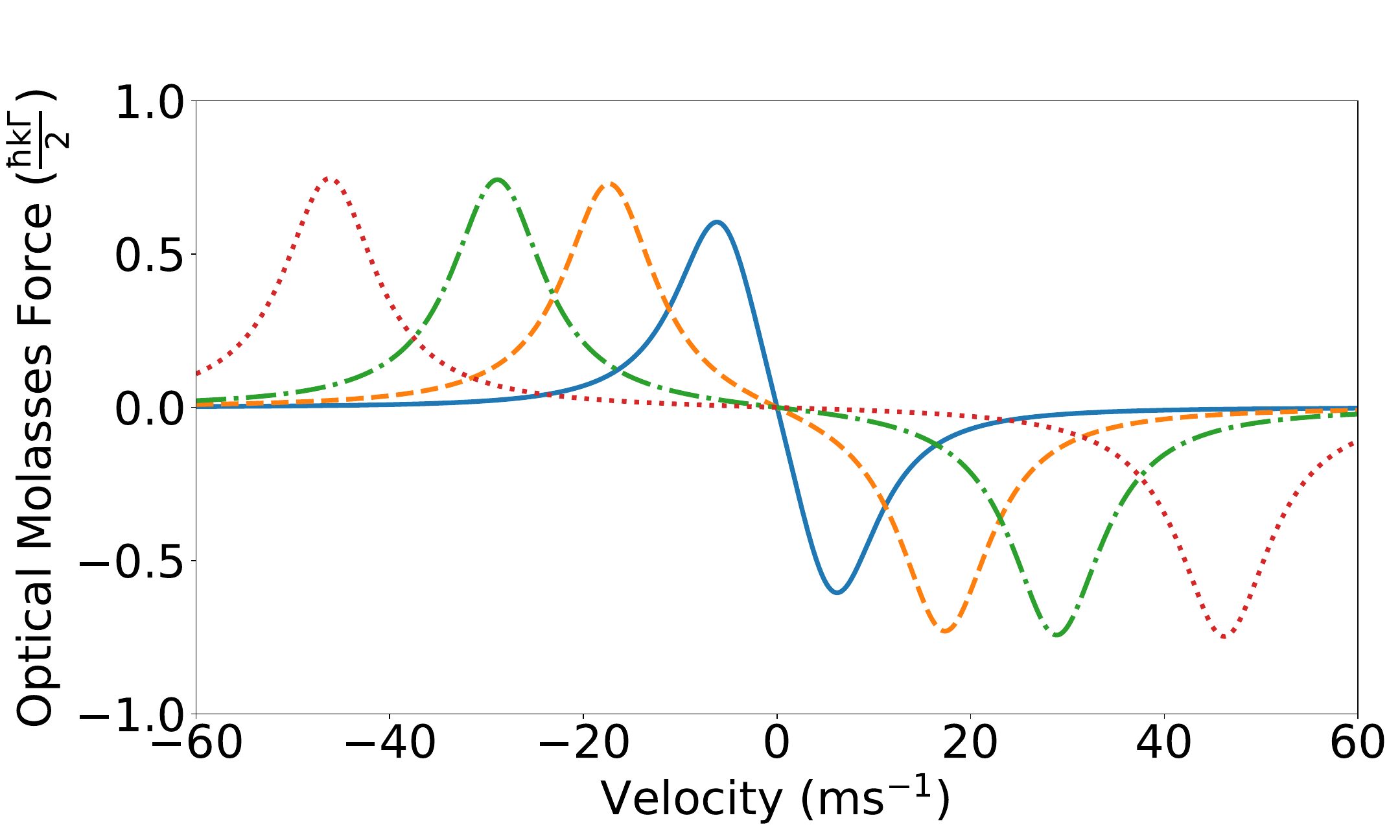}
  \caption{Scattering force against velocity for variable laser detuning, s = 3, $\Delta$: -1 (blue solid), -3 (yellow dashed), -5 (green dash-dot), -8 (dotted).}
  \label{fig:FvsV_var_D}
\end{figure}

Capture velocity ($\mathrm{C_{v}}$) is an important parameter determining the atom number in Eq. (\ref{eq:recaptured_eq}) \cite{bagnato2000measuring, anwar2014revisiting}. Consider an atom starting at the edge of trap an incrementing its initial velocity from $\SI{0}{\metre \second^{-1}}$ until trapping criteria are no longer satisfied. The atom can also never leave the trap radius during a simulation. The highest velocity for which these conditions are met is the capture velocity. Fig. \ref{fig:Na_23_CAPTURE_VELOCITY} replicates work for $\mathrm{{}^{23}Na}$ providing confidence in modelling with $\mathrm{{}^{87}Rb}$ \cite{metcalf1999laser, molenaar1996photoassociative}. A crude estimate for $\mathrm{C_{v}}$ is obtained by considering the work done in slowing an atom as in Eq. (\ref{eq:capture_velocity}): m is the particle mass, $\mathrm{\sigma_{r}}$ is the trap radius and $\mathrm{F_{max}}$ is the maximum scattering force ($\mathrm{\hbar k \Gamma /2}$). Choosing appropriate values for the $\mathrm{{}^{87}Rb}$ $\mathrm{D_{2}}$ line with $\sigma_{r} = \SI{5}{\milli \metre}$ gives $\mathrm{C_{v} \approx \SI{50}{\metre \second^{-1}}}$. This approach assumes a constant maximum scattering force whereas in reality it carries velocity dependence. Assuming  $\Delta = -3$ in the range of $0 - 30 \, \SI{}{\metre \second^{-1}}$ the mean force is $\sim \mathrm{F_{max}/3}$, Fig. \ref{fig:FvsV_var_D}. Accounting for this makes $\mathrm{C_{v} \sim \SI{30}{\metre \second^{-1}}}$. 

\begin{gather}
    \mathrm{C_{v} \simeq \sqrt{\frac{4F_{max}\sigma_{r}}{m}}}.
    \label{eq:capture_velocity}
\end{gather}

To compute $\mathrm{C_{v}}$ more accurately we run the capture velocity simulation at s = 3. 

\begin{figure}[H]
  \centering
  \includegraphics[width=0.47\textwidth, trim={5 1 5 20},clip]{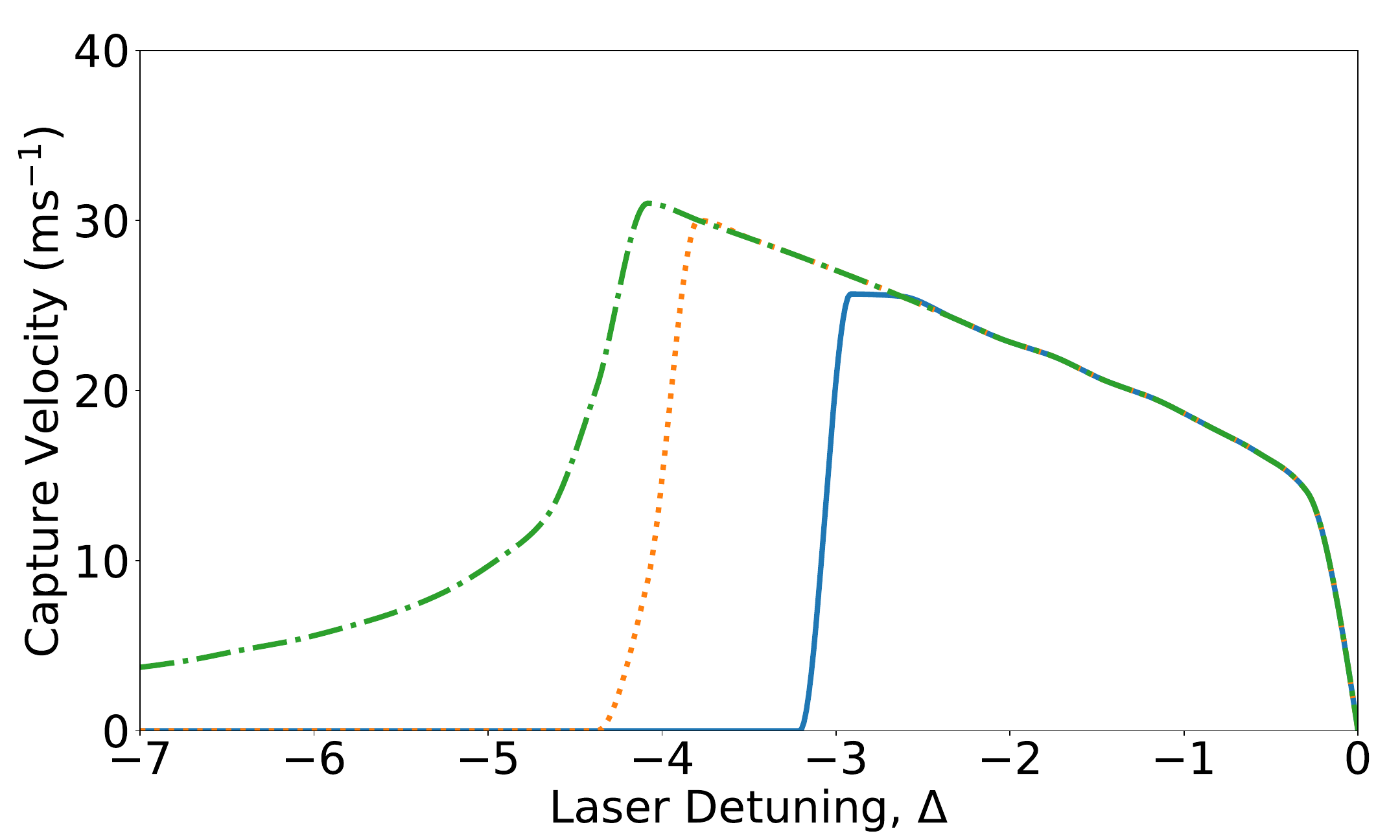}
  \caption{Dependence of capture velocity on simulation time for $\mathrm{{}^{87}Rb}$. A = 14 G/cm, $\mathrm{\sigma_{r}}$ = 5 mm, s = 3, Simulation time (ms) 4 (blue solid), 10 (orange dotted), 50 (green dash-dot).}
  \label{fig:Cv_VARIABLE_SIM_TIME}
\end{figure}

Fig. \ref{fig:Cv_VARIABLE_SIM_TIME} highlights that for increasing simulation time higher velocity atoms can be slowed meaning the peak shifts to greater detunings. Eventually the peak value remains fixed for increasing simulation time with the drop off becoming less extreme. Given our experimental parameters and the short timescale dynamics we adopt $\mathrm{C_{v} = \SI{21}{\metre \second^{-1}}}$ as this provides strong agreement with our MOT loading data and quite good agreement with the simulation value. \\

\end{document}